\documentstyle[psfig]{cupconf}

\ifoldfss
\else
  \ifnfssone
    \newmathalphabet{\mathit}
      \addtoversion{normal}{\mathit}{cmr}{m}{it}
      \addtoversion{bold}{\mathit}{cmr}{bx}{it}
    \newmathalphabet{\mathcal}
      \addtoversion{normal}{\mathcal}{cmsy}{m}{n}
    \else
    \ifnfsstwo
    \fi
  \fi
\fi

\title[Observations of AGN]{Black holes in Active Galactic Nuclei:  observations}
 
\author[G. Madejski]
{G\ls R\ls E\ls G\ns M.\ns M\ls A\ls D\ls E\ls J\ls S\ls K\ls I$^1$}

\affiliation{$^1$Laboratory for High Energy Astrophysics, NASA/Goddard,
Greenbelt, MD 20771\\
[\affilskip] and Dept. of Astronomy, Univ. of Maryland, College Park, MD}

\begin{document}

\ifnfssone
\else
  \ifnfsstwo
  \else
    \ifoldfss
      \let\mathcal\cal
      \let\mathrm\rm
      \let\mathsf\sf
    \fi
  \fi
\fi

\maketitle 

\centerline {\bf To appear in {\sl Theory of Black Hole Accretion
Disks,} }
\centerline {\bf M. Abramowicz, G. Bjornsson, \& J. Pringle, Eds. (Cambridge
University Press)}
\vskip 0.4 cm
\begin{abstract}

This paper summarizes the observations which provide the best 
evidence for the presence of black holes in active galactic
nuclei.  This includes:  X--ray variability; kinematical studies 
using optical emission lines as well as the distribution of megamaser 
spots; and the shape of the Fe K$\alpha$ X--ray emission line.  
It also presents the current status of our understanding of 
jet-dominated active galaxies (blazars), and briefly reviews the 
currently popular AGN ``Unification Schemes'' 
based on orientation effects.  Finally, it reviews the 
observations of the X--ray and $\gamma$--ray continuum, which, 
at least for the radio-quiet objects, is likely to be the 
primary form of their radiative output, and summarizes 
the best current models for the radiative processes responsible 
for the high-energy electromagnetic emission in radio-quiet AGN, 
as well as in jet-dominated blazars.  

\end{abstract}

\firstsection

\section*{1. Introduction}

Perhaps the most exciting astronomical observation leading to our current 
understanding of black holes has been the discovery of quasars.  These 
celestial objects, originally found in the early sixties as point-like 
radio emitters, were identified with apparently stellar sources, possessing 
somewhat unusual spectra, with prominent emission lines.  The identification 
by \cite{schm} of these lines as redshifted systems implied that quasars 
are distant and extremely luminous, commonly producing $10^{46}$ 
erg~s$^{-1}$;  
this is a hundred times or more in excess of the total luminosity of all the 
stars in a galaxy.  Sensitive imaging of the nebulosities which often 
surround them implied that quasars are nuclei of galaxies, and thus are 
higher-luminosity counterparts of the compact nuclei of Seyferts, 
studied some twenty years before the discovery of quasars as unusual emission 
line objects:  hereafter, we assume that they are respectively the lower and 
higher luminosity end of the same population.  A variety of scenarios were 
advanced to explain their nature, and this included multiple supernovae or 
massive spinning stars, but the proposal that quasars are powered by an 
accretion of surrounding matter onto a black hole, advanced in the mid-60s 
by \cite{salp} and \cite{zeld}, 
became the paradigm that we are developing and 
testing today.  While this is a viable and very attractive paradigm, only 
the last few years brought a solid evidence for it, allowing also to measure 
the mass of the central object.  It is important to note here that quasars 
are much more numerous at a redshift $\sim$ 2 than they are locally, meaning 
that a substantial fraction of galaxies must have undergone the quasar phase.  
It is thus likely that many otherwise normal local 
galaxies harbor supermassive black holes, ``dead quasars.''   In fact, 
as we discuss below, there is a number of relatively anonymous galaxies 
that show no signs of nuclear activity, but {\sl do} show evidence for 
such black holes.  

We present the observational evidence for black
holes in AGN in Section 2;  in Section 3, we discuss the effects of
the orientation of the accretion disk surrounding the back hole 
on the appearance of the 
nucleus.  In Section 4, we discuss the jet-dominated AGN known as
blazars.  In Section 5, we review the observations of AGN in X--rays
and $\gamma$--rays, the bands that sample the regions closest to the
black hole, and in Section 6, we review the radiation processes
proposed to explain the emission in these bands.  

\section*{2. Lines of Evidence for Presence of Black Holes in Active 
Galaxies}

There are two general lines of argument that are used to ``prove'' the 
existence of black holes in AGN.  The first attempts to measure the total 
mass within a volume, and argues that no other form besides a black hole 
can have these parameters.  This is done either via estimation of the 
volume from variability data (via the light travel time arguments) and mass 
from the luminosity (via the Eddington limit);  alternatively, this can be 
determined by a measurement of velocity of matter at a specified distance 
from the central object, essentially using Kepler's laws.  The second 
method, discussed in more detail in Chapter 5.3 (by A. Fabian), relies on 
the distortion of the emission line shapes caused by strong gravity 
resulting from the presence of a black hole, and we cover it here only
briefly.  

\subsection*{2.1. X--ray Variability}

The variability of active galaxies generally shows the highest amplitude 
and the shortest time scales in the X--ray and $\gamma$--ray bands,
which happen to be clearly separated from the optical / UV bands by the strong 
absorption of the interstellar medium in our own Galaxy.  This rapid
variability as well as other lines of argument indicate that the 
X--ray / $\gamma$--ray radiation arises the closest to the central
source, and in many cases, is the primary source of energy in
active galaxies.  While the total bolometric luminosity of quasars is
often dominated by the optical and UV flux (see, e. g., Laor et
al. 1997;  for a recent review, see Ulrich, Maraschi, \& Urry 1997), 
the bulk of this flux probably arises in more distant
regions from the central source than the X--rays and $\gamma$--rays.  
The optical and UV radiation arising in the innermost regions 
of the nucleus, on the other hand, is most likely a result of reprocessing of 
X--ray / $\gamma$--ray photons.  This -- as well as the author's 
personal interest in X--rays -- is the reason why this chapter focuses
primarily on the high energy emission from quasars.  

In general, the X--ray variability of quasars is aperiodic.  
While a measurement of periodic variability would give us a clue to the 
circumnuclear environment and thus the nature of the black hole,
besides the ill-fated NGC~6814 (cf. Madejski et al. 1993), 
no strict periodicity was reported for
any AGN.  However, there were two reports of quasi-periodic
variation of flux of active galaxies:  NGC~5548 (Papadakis \& Lawrence
1993) and NGC~4051 (Papadakis \& Lawrence 1995) inferred from the
EXOSAT data, but the quality of the data is only modest, and these still 
need to be confirmed.  More statistically significant is the
quasi-periodic variability of of IRAS~18325-5926 by \cite{iwasawa},
but this is inferred from only a few ($<$ 10) cycles, and requires
confirmation via 
further monitoring before drawing any detailed conclusions.
Nonetheless, we understand relatively little about the details 
of variability of active galaxies, although the recent 
light curves are sufficiently good to discriminate if the time series are 
linear or non-linear;  this is discussed later in this chapter.   

In any case, this rapid variability implies a compact source size.  
This is of course the standard causality argument:  no stationary 
source of isotropic radiation can vary faster than the time it 
takes for light to cross it.  In the X--ray band, the power spectrum 
of variability generally is rather flat at long time scales, and above 
some characteristic frequency, it shows a power-law behavior, such
that the variable power drops with decreasing time scale (see, e.g., 
McHardy 1989);  the Fourier phases of these light curves show no 
coherence (cf. Krolik, Done, \& Madejski 1993).  As it was pointed 
by many authors, for this form of variability, the doubling time 
scale has no definitive meaning, but for the lack of better data, 
it suffices for the illustrative purposes:  it is certainly valid 
as an order-of-magnitude relationship between the source radius 
$r$ and the time scale for doubling of the source flux $\Delta t$ such 
that $r < c$ $\Delta t$.  Again, for quasars, this is particularly 
true in the X--ray band, where the variability is most rapid: even the 
early X--ray data gave us a clue that quasars are very compact.  
For example, an X--ray light curve for the Seyfert galaxy NGC~5506 
(by no means an extreme object) as observed by the ME detector 
onboard the EXOSAT satellite shown on Fig.~1 illustrates it well.  
This, and other 
observations of it, with redshift $z = 0.007$ and a 2 - 10 keV X--ray flux 
of $\sim 4 \times 10^{-11}$ erg cm$^{-2}$ s$^{-1}$, imply a luminosity 
$L_{\rm X}$ of $\sim 10^{43}$ erg s$^{-1}$, corresponding to an Eddington 
mass of at least $\sim 10^5$ $M_{\rm o}$.  This corresponds to a Schwarzschild 
radius $r_{\rm S}$ of $3 \times 10^{10}$ cm.  The X--ray data show a doubling 
time of $\sim 10,000$ s, corresponding to $r < 3 \times 10^{14}$ cm, which, 
for a $10^5$ $M_{\rm o}$ black hole, would imply that the X--ray emission 
arises from a region of radius $r_{\rm X} \sim 10^4$ $r_{\rm S}$.
This example is 
by no means extreme -- we made a number of assumptions that are 
probably even too conservative:  more realistic assumptions 
imply a mass of $10^6$ $M_{\rm o}$, and $r_{\rm X} \sim 10^3$ $r_{\rm S}$.  
A number of more extreme cases -- including very luminous objects -- were 
reported recently on the basis of the ROSAT data by 
\cite{boller}.  These are generally for the so-called 
``narrow-line 
Seyfert~1s,'' and we will return to those objects later.  In brief, a 
doubling time scale of $\sim 1000$ s for a source with $L_{\rm X} \sim 
10^{44}$ erg s$^{-1}$ is not uncommon, implying that the bulk of the 
X--ray emission arises around 10 - 100 $r_{\rm S}$.  However, the use 
of the observed variability time scale 
does not provide an ``airtight'' argument for the size of the emitting 
region, since the observed emission may well be anisotropic, yielding
an underestimate of the emitting volume, as is almost certainly the 
case for blazars.  We discuss this in more detail later on.  

\begin{figure} 
\vspace {-60pt}
\centerline{\psfig{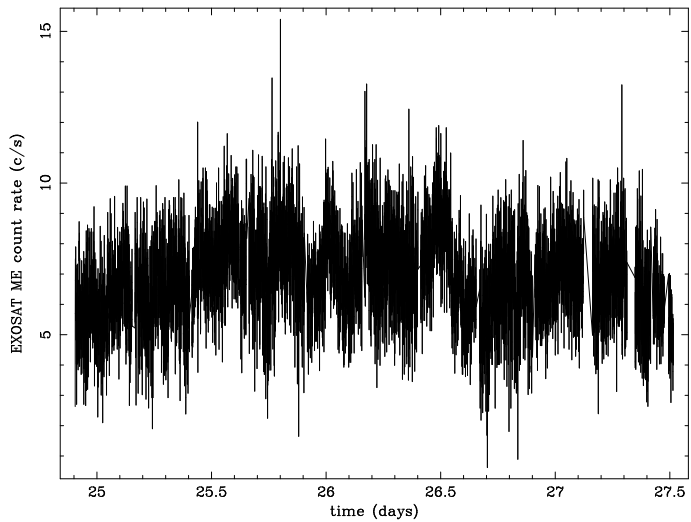}}
\vspace{10pt}
\caption{
X--ray light curve for the Seyfert galaxy NGC~5506, collected with the
EXOSAT ME satellite
(from Krolik, Done, \& Madejski 1993).  
}\label{5506}
\end{figure}

\subsection*{2.2. Kinematic Studies of Active and ``Normal'' Galaxies 
Using Optical Emission Lines}

The other line of evidence for the presence of black holes in galaxies 
(both active and ``normal'') is the velocity field of the matter emitting 
closely to the nucleus.  This kind of work has been recently reviewed by 
\cite{korm}, and it dates back to the ground-based 
observations made in the late 70s, when W. Sargent and collaborators showed 
that the stellar velocity dispersion in the radio galaxy M~87 increases to 
350 km~s$^{-1}$ in the innermost 1.5$''$ from the nucleus.  M~87 was in fact 
observed by the Planetary Camera by \cite{ford87}
and the Faint Object Spectrograph by \cite{harms} 
onboard the repaired Hubble Space Telescope, and 
the images showed the presence of a disk-like structure of ionized gas 
in the innermost few arc seconds.  The spectroscopy provided a measure 
of the velocity of the gas at an angular distance from the nucleus of 
0.25$''$ (corresponding to $\sim 20$ pc, or $\sim 6 \times 10^{19}$ cm), 
showing that in the reference frame of the object, it recedes from us on 
one side, and approaches us on the other, with a velocity difference of 
$\sim 920$ km s$^{-1}$ (see Fig.~2).  This implies a mass of the central 
object of $\sim 3 \times 10^{9}$ $M{\rm_o}$, and besides a black hole,
we know of no other form of mass concentration that can ``fit'' inside
this region.  
As an aside, it is worth noting that M~87 is known to have a relativistic jet 
perpendicular to the disk structure mentioned above, expanding with 
the bulk Lorentz factor $\Gamma_j$ of 4.  As such, this object is probably 
just a blazar, with the jet oriented at an angle $\sim 40^{\rm o}$ to 
the line of sight, and thus is probably the closest to a hard evidence 
that blazars (which we discuss in more detail below) 
indeed {\sl do} harbor black holes.  

\begin{figure} 
\centerline{\psfig{file=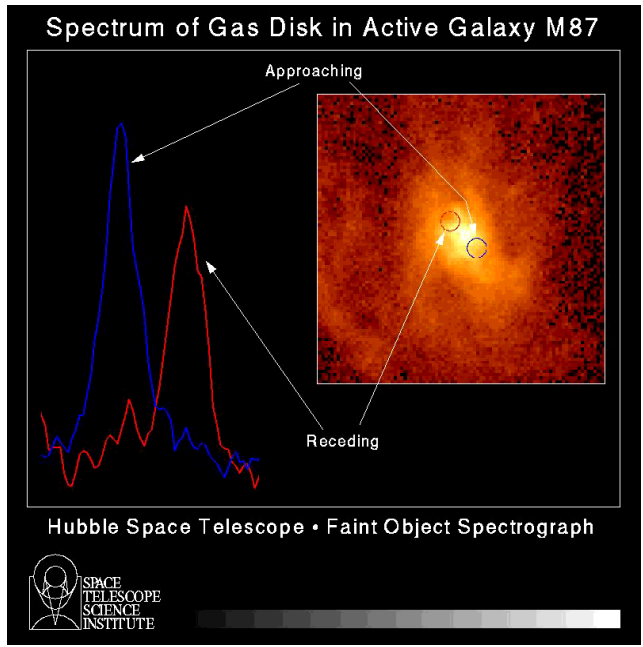,height=4.3in}}
\vspace{-40pt}
\caption{
Illustration of the spectra of gas in the vicinity of the nucleus of
the radio galaxy M~87 (STScI Public Archive).  
}\label{m87}
\end{figure}

Besides M~87, similar spectroscopy observations done with Hubble Space 
Telescope revealed high stellar velocities in the central regions of a 
number of normal galaxies which otherwise show no evidence for an active 
nucleus.  These were summarized recently by \cite{ford}, and include 
well-publicized observations (by L.~Ferrarese, H.~Ford, J.~Kormendy and 
others) of NGC~6251, NGC~4261, NGC~4594, NGC~3115, but doubtless by
now probably there are several new objects.  Such high velocities in 
the innermost regions of these galaxies cannot be explained in any 
other way besides invoking the presence of massive ($10^{8} - 10^{9}$ 
$M_{\rm o}$) black holes in their centers.  (Interestingly, several of
these galaxies also show weak radio jets!)  
Even for our own Milky Way galaxy, the infrared data and velocity 
measurements -- readily performed from the ground (cf. Eckart \& Genzel
1997), at much higher resolution  
than the HST data for external galaxies -- 
reveal a ``modest'' nuclear black hole with a mass of $\sim 3 \times 10^{6}$ 
$M_{\rm o}$.  The evidence is building that supermassive black holes are 
quite common;  recent estimates by \cite{ford} as well as \cite{ho} 
imply that they 
inhabit perhaps as many as half of all galaxies, and may well be the 
``dead quasars'' of the past.  Active galactic nuclei are thus probably 
only the ``tip of the iceberg'' of their population.  

\subsection*{2.3. Megamasers in Active Galaxies}

Perhaps the most elegant observation showing the presence of a Keplerian 
disk around a black hole -- and thus capable of measuring the mass of the 
hole independently of the otherwise uncertain estimates of its distance -- 
was the Very Large Baseline Interferometry observation of megamasers in the 
vicinity of the nucleus of Seyfert~2 galaxy NGC~4258 reported by 
\cite{miyoshi}.  The masing activity can only be observed 
along a line of sight where the velocity gradient is zero, meaning that these 
can be seen at locations with masers either between us and their 
source of energy, or locations at $\sim$ 90$^{\rm o}$ to the line of sight.  
Indeed, this is the spatial distribution of maser spots around the nucleus 
of NGC~4258, illustrated in Fig.~3.  Specifically, these observations reveal 
individual masing spots revolving at distances ranging from $\sim 0.13$ pc to
about twice that around the central 
object -- presumably, again, a black hole -- with a mass of $\sim 3.6 \times 
10^{7}$ $M_{\rm o}$.  What is truly remarkable about these data is the 
near-perfect Keplerian velocity distribution, in a slightly warped disk-like 
formation;  this implies that almost all the mass is located well within the 
inner radius where the megamasers reside.  It is not possible to have a 
cluster of distinct, dark, massive objects responsible for such gravitational 
potential;  at least some of the objects would escape on a relatively short 
time scale, and form a potential well with a different shape, which would 
now force a departure of the megamaser--emitting material from pure 
Keplerian motion (cf. Maoz 1995).  

\begin{figure} 
\centerline{\psfig{file=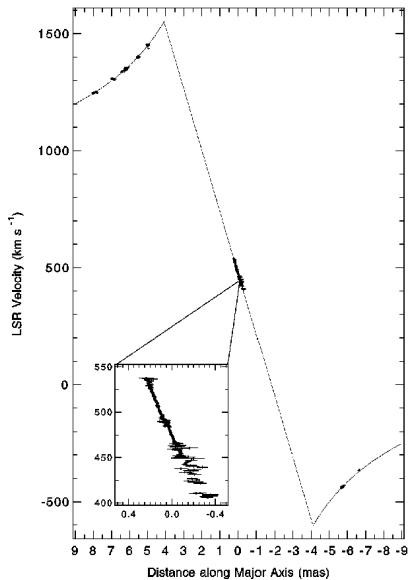,bbllx=0.5in,bblly=2.5in,height=3.0in}}
\vspace{10pt}
\caption{
The distribution of megamaser spots in the Seyfert~2 galaxy
NGC~4258
(from Miyoshi et al. 1995).  
}\label{4258}
\vspace{-1pt}
\end{figure}

Since we know the central mass quite precisely, NGC~4258 has been a terrific 
laboratory to study the details of the accretion disk.  In particular, this 
is a relatively low luminosity ($\sim 10^{42}$ erg s$^{-1}$) object, making 
it quite sub-Eddington, with $L/L_{\rm E} \sim 3 \times 10^{-4}$.
Such sub-Eddington sources are likely to obey unique solutions of
accretion disk structure (see, e.g., Ichimaru 1977;  Narayan \& Yi
1994;  Abramowicz et al. 1995), where the accreting gas is optically
thin and radiates inefficiently, and the accretion energy that is dissipated
viscously, is advected with the accretion flow.  With this, as was argued by
\cite{lasota}, the accretion disk in NGC~4258 can well 
be advection-dominated (but this does not {\sl have} to be the case;
see, e.g., Neufeld \& Maloney 1995).  However, it is important to note 
that such low 
Eddington rate cannot be universal among quasars;  if most of them radiated 
at such low $L/L_{\rm E}$, the black hole masses of the most luminous sources 
would be much larger than expected on other grounds.  However, an
intriguing possibility (cf. Yi 1996) 
is that quasars in their ``youth,'' when the
black hole masses were more modest, had standard, ``cold'' (Shakura -
Sunyaev) accretion disks, and this is why they were so luminous in the
past.  The masses of the black holes grew with time, and even if the
mass rate supplied for accretion remained constant, $L/L_{\rm E}$ 
actually decreased, and thus the the inner accretion disks switched from
``cold'' (bright, Shakura - Sunyaev) phase to ``hot'' 
(fainter, advection-dominated) phase even if the rate of mass supply {\sl did
not} decrease.  This is perhaps why some of the yesterday's bright 
quasars are today's dormant, ``dark'' black holes, revealed only via the 
kinematical studies mentioned above.  

Nonetheless, NGC~4258 is probably {\sl not} a unique 
object.  We know of a large class of ``low activity'' active galaxies, known 
collectively as ``Low Ionization Nuclear Emission Region'' objects, or 
LINERs;  they generally have low luminosity, coupled with the absence of 
the luminous inner disk as evidenced by the emission line ratios.  However, 
unlike the {\sl bona fide} Seyferts with low luminosity, which seem to vary 
relatively rapidly in X--rays, implying they are ``scaled down'' quasars 
with relatively low mass black holes -- LINERs are known {\sl not} to vary 
rapidly in any band, suggesting that the low activity is not due to a low 
black hole mass, but rather due to a low accretion rate (cf. Ptak 1997).  
As the nuclei of 
these objects are not very bright, the data are sparse, and thus the details 
of the radiative processes are poorly known;  while workable models
exist (see, e.g., Lasota et al. 1996), 
they still require more work on the details of the transition between the 
``standard'' and advection-dominated regions of the disk.  

\subsection*{2.4. Profile of the Fe K Emission Line} 
Perhaps the most convincing evidence that a strong gravitational field 
is present in active galactic nuclei comes from the recent measurements 
of the shape of the Fe~K$\alpha$ fluorescence line, arising in a geometrically 
thin, but optically thick accretion disk.  This is discussed in more
detail by A. Fabian (Chapter 5.3), so what follows is a brief summary.  
The inner part of the disk 
is illuminated by X--rays.  Because of the relative Cosmic abundances 
and the fluorescence yields of various elements, the strongest discrete 
spectral feature predicted from the disk is the 6.4~keV fluorescent 
Fe~K$\alpha$ line;  the strength (equivalent width) of the line of
$\sim$ 150 eV, as measured by \cite{pounds}, 
is in fact roughly consistent with predictions of \cite{george}.  Since 
this line arises from matter in motion, its profile is a tracer of the 
velocity field of the accreting matter.  

\begin{figure} 
\vspace{-70pt}
\centerline{\psfig{file=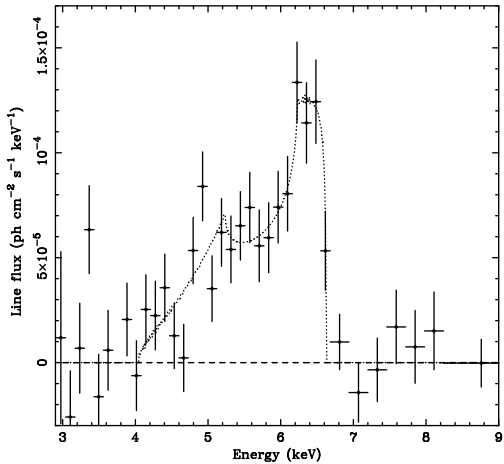,height=3.7in}}
\vspace{0pt}
\caption{
The shape of the Fe~K$\alpha$ fluorescence line, observed in the
Seyfert galaxy MCG-6-30-15, showing the characteristic two-pronged
shape expected when the emitting matter is in a disk-like structure
orbiting closely to a black hole  
(from Tanaka et al. 1995).  
}\label{mcg}
\vspace{-10pt}
\end{figure}

The Asca observations of the X--ray bright Seyfert galaxy MCG-6-30-15 
by Tanaka et al. (1995) indeed showed that the line (see
Fig.~4) has a characteristic two-pronged shape expected to arise from 
matter flowing in a disk-like structure.  The matter approaching us 
is responsible for the blue wing of the line, while that receding 
produces the red wing, with an additional redshift, since the photons 
are emitted in a strong gravitational field;  the exact shape also 
depends on the inclination of the disk.  A detailed spectral fitting 
of the line shape indicates that the emitted energy of the line is 
indeed 6.4 keV, while the bulk of its flux arises at $<$~10~$r_{\rm S}$, 
implying in turn the presence of nearly neutral material very closely 
to the black hole.  The analysis of a number of Seyfert spectra from 
the Asca archives by \cite{nandraline} suggests that many 
Seyferts indeed show the Fe K$\alpha$ line profiles that require an 
emission close to the black hole, but the quality of data is only 
modest.  Fortunately, a number of more sensitive observatories -- 
such as AXAF, Astro-E, XMM, and Constellation-X, 
will be launched in the next few years,
providing ample opportunities for X--ray observations of effects of 
strong gravity. 

\section*{3. Unifying Seyfert 1s and Seyfert 2s:  the Orientation Effects}

The megamaser source NGC~4258 is only the first of three active 
galaxies showing a spatial distribution of masing spots from which 
it is possible to measure the mass of the central object.  The other 
two are the well-known NGC~1068 (Greenhill 1998), and NGC~4945 
(Greenhill, Moran, \& Herrnstein 1997).  Both are 
also classified as Seyfert~2s, which from the observational 
side means that they show narrow emission lines, implying 
velocities on the order of 1,000 km~s$^{-1}$;  these lines show no 
variability, and thus it is generally accepted that they originate in a 
relatively large regions, on the order of 100~pc or more.  Seyfert~1s as 
well as luminous quasars, on the other hand, exhibit generally very 
different spectra, with permitted emission lines, and these lines are 
usually broad, with velocities upwards of 1,500 km~s$^{-1}$, often reaching 
30,000 km~s$^{-1}$.  This, together with the variability of the lines that 
is commonly observed on time scales of weeks or months, implies that the 
broad line region is located much closer to the nucleus than the
narrow line region.  

The likely relationship between the two classes of active galaxies was 
revealed by the seminal observation of the well-known Seyfert~2 galaxy 
NGC~1068 by Antonucci \& Miller (1985).  The spectropolarimetric 
study of the H$\beta$ line revealed that when observed in polarized light, 
the line is broad.  They interpreted the polarization as due to electron 
scattering by material that is distributed preferentially 
along the symmetry axis of the system, and advanced the widely accepted 
scenario explaining the differences between the two types of Seyferts as an 
orientation effect.  This is illustrated in Fig.~5;  all Seyfert galaxies 
are surrounded by a geometrically and optically thick torus, with an inner 
radius of a fraction of a parsec.  Such a torus can be, for instance,
the outer regions of a 
severely warped disk, as recently suggested by M. Begelman and
J. Pringle (see Chapter 8.1).  
When an object is viewed along the axis of the torus, it is a Seyfert~1, 
revealing all the ingredients of the nucleus:  the broad line region and 
the unobscured X--ray source, both commonly varying on a short time scale.  
When viewed in or close to the plane of the torus, the broad line region 
is completely obscured, and the soft end of the X--ray spectrum is absorbed 
due to the photoelectric absorption by the material in the torus.  The 
opening of the torus contains a ``mirror'' of ionized gas with free 
electrons, and these are responsible for the scattering of the broad 
line light into the line of sight, and hence the broad lines are seen 
only in polarized light.  

\begin{figure} 
\centerline{\psfig{file=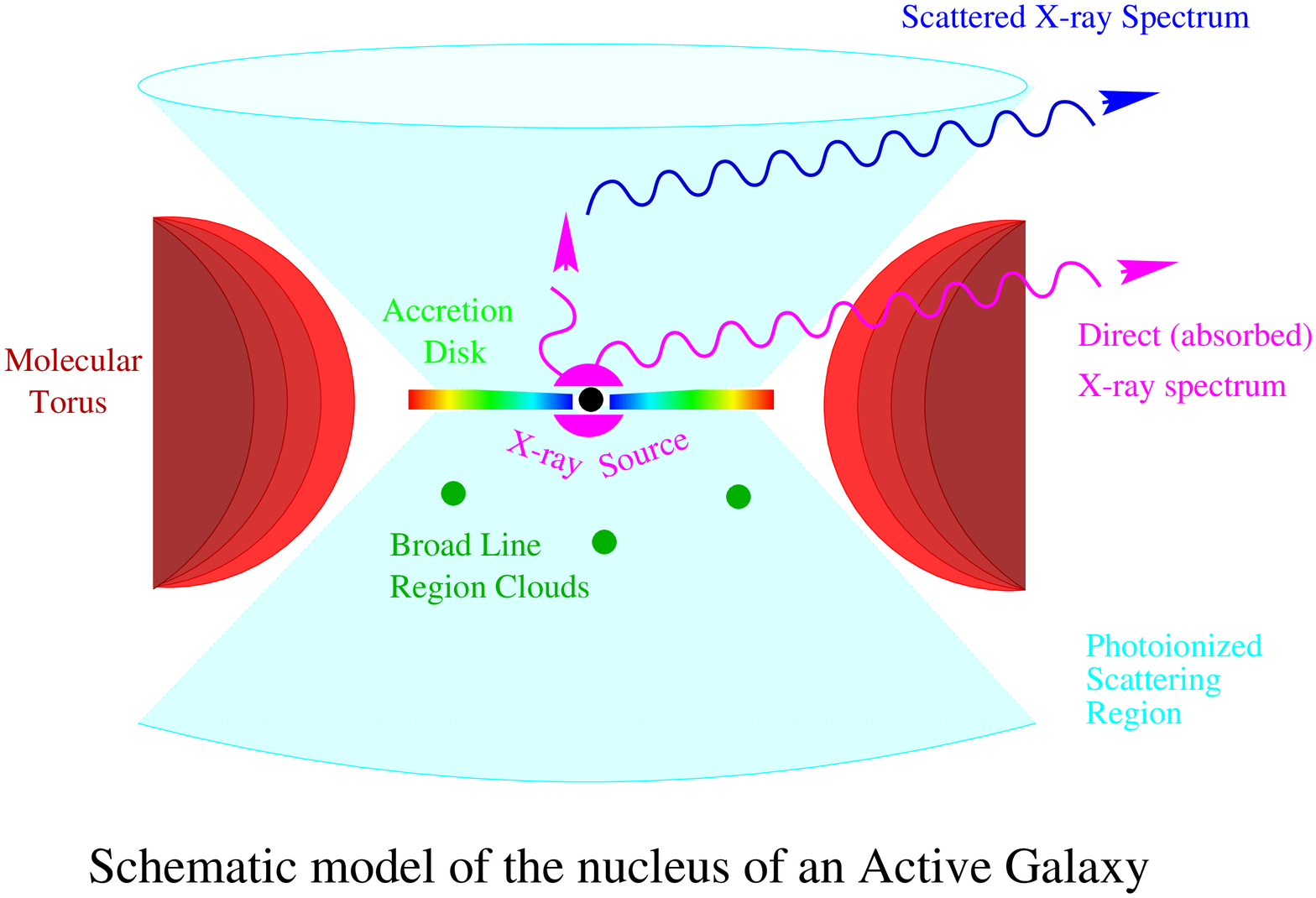,height=2.7in}}
\vspace{5pt}
\caption{
Schematic illustration of the ``Unified Picture'' of Active
Galactic Nuclei:  when viewed along or near the axis of the 
torus/disk, it is
a Seyfert~1, while viewed along the plane, it is a Seyfert~2 
(courtesy of Dr. C. Done, Univ. of Durham).  
}\label{Unif}
\vspace{-15pt}
\end{figure}

In the context of this ``unification'' model, it is thus not surprising that 
all three megamasers with measured rotation curves are Seyfert~2 objects:  
to even observe a megamaser, we have to be located in the plane of material 
which has large column density and presumably is roughly co-axial with 
the torus.  Likewise, the X--rays traveling closely to the plane of the 
torus encounter a larger column density, resulting in the greater 
photoelectric absorption.  This is in fact, observed;  even early X--ray 
spectra of Seyferts showed preferentially larger absorption in sources 
showing only narrow lines (Seyfert~2s), with the broad line region obscured 
by the same material that absorbs soft X--rays.  Note that this is 
different than the case of only narrow emission lines present in some
low-luminosity radio galaxies, such as M~87 (collectively known as the FR-I 
objects):  there, the broad lines cannot be obscured, as we see no 
evidence of X--ray absorption.  Furthermore, while Seyfert 2s
sometimes do show highly ionized, narrow permitted lines (in addition
to the commonly detected forbidden lines), FR-I objects {\sl do not}, 
implying an absence of the very strong isotropic ionizing UV
continuum.  We will return to this point below.   

\section*{4. Anisotropy and Doppler Effects:  the Case of Blazars}

The use of the observed variability time scale does not, however, provide 
an ``airtight'' argument for the size of the emitting region, since the 
observed emission may well be anisotropic, yielding an underestimate of the 
emitting volume.  As pointed out by \cite{rees}, 
if the radiating 
matter moves at a relativistic speed towards the observer, the variability 
time scale appears shortened by the Doppler effect.  Furthermore, the flux 
of radiation is boosted in the direction of motion of the emitting matter, 
and thus the total luminosity inferred under an assumption of isotropy is 
an overestimate.  We are nearly certain that this is the case for 
blazars.  This sub-class of quasars consists of sources 
that generally show strong radio emission and 
are highly variable and polarized in all observable wavelengths, and 
also includes members with very weak or absent emission lines, known as 
BL Lacertae -- type objects.  High angular resolution radio observations, 
performed since the late 60s -- using Very Long Baseline Interferometry -- 
implied compact radio emission regions, often associated with a rapid 
change of their structure.  With the assumption that the redshifts are 
indeed cosmological, these structures appeared to expand at transverse 
speeds exceeding $c$, and in many cases, they had shapes of jets.  If the 
speed of the emitting matter is very relativistic, pointing closely to our 
line of sight, the apparent superluminal motion is just a projection effect, 
providing further support for the Doppler-boosting scenario for blazars.  

In a number of aspects, blazars are the most extremely 
active galactic nuclei.  Observations of many blazars by the 
EGRET instrument onboard the Compton Gamma-ray Observatory (CGRO) 
indicated that usually they are strong emitters in 
the GeV $\gamma$--ray band;  we know of at least 50 such objects.  The GeV 
emission can dominate the overall observed electromagnetic output of these
sources, and this 
is illustrated in Fig.~6 for 3C 279, the first of the ERGET-discovered 
GeV blazars.  In a few cases, the $\gamma$--ray emission has been observed 
with Cerenkov radiation telescopes to extend as far as the TeV range.  This 
is particularly exciting, as TeV radiation can be observed from the ground, 
allowing a study of quite exotic phenomena occurring in quasars without 
the expense of space-borne platforms.  Interestingly, we know of no 
``radio-quiet'' MeV or GeV $\gamma$--ray emitting quasars, 
implying an association of the $\gamma$--rays with the compact 
radio source and thus a jet.  

\begin{figure} 
\vspace{-40pt}
\centerline{\psfig{file=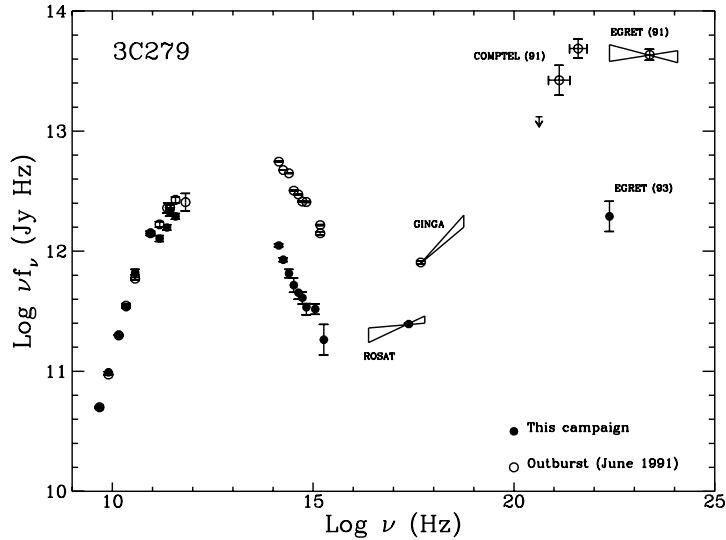,height=3.0in}}
\vspace{40pt}
\caption{
Broad-band spectrum of the GeV-emitting blazar 3C279, 
illustrating the dominance of the GeV $\gamma$--ray emission over
other bands
(from Maraschi et al. 1994).  
}\label{3c279}
\end{figure}

In many cases, the $\gamma$--ray emission from blazars is variable 
on a time scale of a day, and recent, simultaneous multi-wavelength 
monitoring observations indicate that the flux swings are reasonably well 
correlated between various bands, with the amplitude of variability 
increasing with the increase of the energy of the observing band.  If the 
emission in the lower energy bands arises from a region co-spatial with that 
where the $\gamma$--rays originate -- and the tracking of light curves in 
various bands supports this -- then the opacity to the $\gamma$-$\gamma$ 
pair production would imply large optical thickness.  Similarly, even the 
early calculations indicated that if the radio sources are as luminous and 
as compact as the variability data imply, the cooling should be 
predominantly via the Compton process, with the radiating electrons losing 
all their energy via upscattering of the just--produced--photons, 
regardless of the radiation mechanism.  Even the large $\gamma$--ray 
fluxes detected by EGRET are often orders of magnitude lower than 
the predictions of this self-Compton process.  Both these difficulties, 
however, go away if the radio as well as the GeV emission are 
both Doppler-boosted, meaning that the true luminosity is lower, 
and source sizes are greater than inferred under an assumption of 
isotropy.  In fact, we now believe that the entire broad-band 
continuum is produced in the relativistic jet (cf. K\"onigl 1981;  
Ulrich et al. 1997), and statistical 
considerations imply that the bulk Lorentz factors $\Gamma_j$ 
of these jets are on the order of 10 (cf. Vermeulen \& Cohen 1994).  

For the case of BL Lac -- type blazars, the Doppler enhancement (going 
roughly as a square of $\Gamma_j$;  see, e.g., Appendix A of Sikora et
al. 1997) ) is so strong that the continuum 
completely outshines the emission lines;  in some cases, however, 
it is possible that the emission lines are absent to begin with.  
In any case, the mechanism for an acceleration and collimation of these 
powerful jets to such high relativistic speeds is not known, but general 
considerations of jet energetics are strongly suggestive that this energy 
is ultimately tapped from an accretion onto a compact object:  the 
best picture today involves the so-called Blandford-Znajek mechanism 
(Blandford \& Znajek 1977), 
where the power of the jet derives from the spin of the black hole.  

\subsection*{4.1. Emission Processes in Blazars}

The electromagnetic emission from jets in blazars -- recently reviewed by 
\cite{sikora} and Ulrich et al. (1997) 
-- is interesting in its 
own right, and the comparison of the broad-band spectra of blazars and 
the ``radio-quiet'' quasars shows it clearly.  The strong GeV component 
which clearly dominates the spectra of the former (cf. Fig.~6) is entirely 
absent in the latter, and the relative strength of the GeV emission as 
compared to that in keV range is often different by a factor of 
$\sim 100$.  The radio and optical emission shows polarization, and spectra 
are always non-thermal.  All these facts argue that the radio-through-UV 
emission is produced by the synchrotron process, with ultrarelativistic 
electrons (with $\gamma_{\rm el} \sim 1000$ or more) radiating in 
magnetic field on the order of a Gauss.  The Compton upscattering of lower 
energy photons mentioned above is indeed present - and most current models 
suggest that it is responsible for the radiation observed above a few keV.  
In the lineless BL Lac type objects, we have no evidence for any strong 
external radiation field, and the dominant ``seed'' photons for Compton 
upscattering are likely to be produced by the synchrotron process internally 
to the jet.  As was shown by \cite{dermer} as well as by \cite{sikbegrees} 
and \cite{blandlev}, in blazars with emission lines, the external, diffuse 
radiation dominates;  since this radiation is isotropic in the stationary 
frame, it appears Doppler-boosted in the frame of the jet.  

This scenario predicts that there should be many objects with jets 
pointing farther away, up to the right angle to our line of sight.  
We now believe that for the quasar-type blazars, with strong emission 
lines, these are the giant, powerful radio galaxies.  The difference 
between the quasar-type and BL Lac type blazars may well be related to 
the presence of the inner, extremely luminous accretion disk in the 
former.  As suggested by M. Begelman, and, more recently \cite{reynolds}, 
some radio galaxies (as, for instance, M~87) have 
massive black holes, yet relatively modest luminosity, and essentially no 
broad emission lines;  this might argue for a very sub-Eddington 
accretion rate, with a likelihood of an advection-dominated inner accretion 
disk.  An intriguing possibility is that these low-luminosity radio 
galaxies are the ``misdirected'' BL-Lac - type blazars:  the absence of 
the inner, luminous disk would then be responsible for the absence of the 
strong ionizing radiation and thus broad emission lines.  This in turn would 
mean that in BL Lac - type blazars, there are no external photons to act 
as ``seeds'' to be Comptonized by the relativistic electrons in the jet, 
and the only ``seed'' photons are internal, produced by the synchrotron 
process.  

\subsection*{4.2. Origin of the Difference Between ``Radio-Quiet'' Active 
Galaxies and Blazars}

We are nearly certain that there is a clear {\sl intrinsic} 
distinction between the 
jet-dominated blazars, and the more common, radio-quiet quasars.  However, the 
reason for the different behavior of these two subclasses is far from 
certain, and this is primarily because we do not have a good understanding of 
the formation and acceleration of relativistic jets.  Even though we 
cannot use the causality arguments to infer the presence of black holes 
in blazars solely from their variability -- since we do not know for sure 
as to what extent the true variability time scales are shortened by 
Doppler boosting -- more indirect evidence implies that ultimately, accretion 
onto a black hole powers blazars as well.  As to the difference
between the two categories, perhaps the most appealing 
scenario, advanced by \cite{wilson}, as well as by \cite {moder}, is 
where the two classes differ by strong or weak spin of the 
black hole;  again, the jet is formed and powered by tapping the rotational 
energy of the hole.  This is somewhat similar to the distinction between the 
Galactic binary ``microquasars'' suggested by \cite{zhang}, 
discussed in Chapter 2.1 (by P. Charles).  Nonetheless, this is an area 
of very active research where no clear conclusions have been reached.  
In particular, if the shape of the fluorescence Fe K line seen in the 
Seyfert galaxy MCG-6-30-15 (see above) and possibly
in other Seyferts (Nandra et al. 1997a)  indeed requires 
that the cold material powering the nucleus flows via a disk-like 
structure, this implies that the fluorescence occurs at at a distance 
$r < 3$~$r_{\rm S}$.  Since the last stable orbit in a non-rotating 
(Schwarzschild) black hole is at 3 $r_{\rm S}$ and beyond this, the matter 
is in a free fall, this would imply that the black hole is spinning (Kerr), 
where the last stable orbit can be substantially closer to the black
hole, depending on its spin.  However, we note that an alternative
scenario, advanced by \cite{reynbeg} does {\sl not} require a disk at 
$r < 3$~$r_{\rm S}$;  nonetheless, this still requires X--ray emission at 
least at or just beyond 3 $r_{\rm S}$, 
merely eliminating the requirement of {\sl spinning} black hole, but it 
still implies that at least a non-rotating black hole is present.  

\section*{5. Zooming in on a Supermassive Black Hole:  High Energy 
Spectra of Radio-Quiet Active Galaxies}

As we mentioned above, active galaxies are generally variable, and the 
most rapid variability is observed in the X--ray and 
$\gamma$--ray bands;  since we are interested in understanding the regions 
closest to the black hole, these bands deserve the most detailed
study.  A recent, excellent article by Mushotzky, Done, \& Pounds (1993)
reviewed the X--ray spectra of AGN;  however, this covered the status of
observations before the results from CGRO became available, while we
do include these results here.  
The most conclusive results are gleaned by a study of the non-blazar 
active galaxies, to avoid any potential contamination by the jet.  
Since no radio-quiet (= jet-less) active galaxy was detected 
above several hundred keV, the most relevant spectral region is the 
X--ray and soft $\gamma$--ray bands.  This covers nearly four decades in 
energy, from $\sim 0.2$ keV to $\sim 1$ MeV, and often requires 
observations with multiple satellites, which must be made simultaneously, 
since active galaxies are variable in all bands.  

The early X--ray observations of active galaxies by \cite{mush}, 
\cite{halp}, and \cite{roth} implied that, to the 
first order, X--ray spectra of active galaxies are power laws with the energy 
index $\alpha$ (defined such that the flux density $S_{\nu} \propto 
\nu^{-\alpha}$) of about 0.7, modified by photoelectric absorption at the 
low energy end.  A major advance came from sensitive observations 
with the ROSAT and Ginga satellites (covering respectively the bands 
of 0.1 to 2, and 2 to $\sim$ 30 keV), revealing that the spectra are more 
complex, with a somewhat softer underlying continuum, with $\alpha \sim 0.9$.  
This continuum is modified by photoelectric absorption in the host galaxy 
of the quasar;  the absorbing material can be either cold, or partially 
ionized, and this manifests itself as isolated edges, most notably of 
oxygen, seen in the ROSAT (and more recently, also in Asca) data.  
The Ginga spectra, reported by 
\cite{pounds}, showed a strong emission line at the 
rest energy of $\sim 6.4$ keV, presumably due to fluorescence of the K 
shell of iron, and a hardening above $\sim 8$ keV.  These last two features 
were interpreted as signatures of reprocessing (``reflection'') of the 
primary continuum in the cold material that is accreting onto the nucleus.  
The Fe K line is indeed the strongest expected, via the combination of the 
relatively high abundance of iron (decreasing with the atomic number $Z$), 
and fluorescence yield (increasing with $Z$);  as pointed out by
\cite{makis}, 
the line equivalent width as measured by Ginga is too high to be produced 
in absorbing material of any column.  

Reprocessing by the matter that accretes onto the black hole is thus the 
most viable alternative.  The hardening, predicted earlier in a seminal 
paper by Lightman \& White (1988), arises as an additional 
spectral component, due to Compton reflection from cold matter.  This is 
produced roughly at one Thomson depth, or when the equivalent hydrogen 
column density is $\sim 1.6 \times 10^{24}$ cm$^{-2}$.  The cosmic 
abundances of elements are such that at low energies, the photoelectric 
absorption dominates, and unless the accreting matter is substantially 
ionized, this component emerges only above the last significant absorption 
edge, again from iron.  The strengths of the reflection 
component and the line are in fact in agreement with theoretical predictions 
by \cite {george}.  With the intensity of the 
incident X--rays 
greatest closely to the central source (as inferred from the variability 
of the continuum), its kinematic and gravitational Doppler shifts are 
powerful diagnostics of the immediate circumnuclear region.  

\subsection*{5.1. A Working Template:  Spectrum of Seyfert 1 IC~4329a}

Perhaps the brightest {\sl bona fide} Seyfert~1 on the sky is the luminous 
object IC~4329a, and it has been observed simultaneously with the ROSAT 
(0.2 - 2 keV) and CGRO OSSE (50 - 1000 keV) (Fabian et al. 1993;
Madejski et al. 1995).  This left a gap between 
$\sim 2$ and $\sim 50$ keV, so the observations were supplemented by 
non-simultaneous data obtained by Ginga a few years earlier;  the Ginga 
data were renormalized to match the ROSAT flux at 2 keV, so this combined 
data set provided a good representation of a broad band high energy 
spectrum of a Seyfert~1, and is illustrated in Fig.~7.  The data indeed 
showed a modest photoelectric absorption due to neutral material (most 
likely the ISM of the host galaxy), plus an additional modest column of 
ionized absorber.  Beyond $\sim 2$ keV, the spectrum is the primary 
continuum, with $\alpha \sim 0.9$.  At $\sim$ 6.4 keV, there is a strong 
Fe~K line, which in the case of IC~4329a and many other Seyfert galaxies 
is broad, with $\sigma$ of at least 200 eV.  The detailed studies 
of this line in another object, MCG-6-30-15 \cite{tanaka}, show the 
characteristic two-pronged shape, implies relativistic motion in an 
accretion disk inclined to the line of sight as discussed above.  

\begin{figure} 
\vspace{-30pt}
\centerline{\psfig{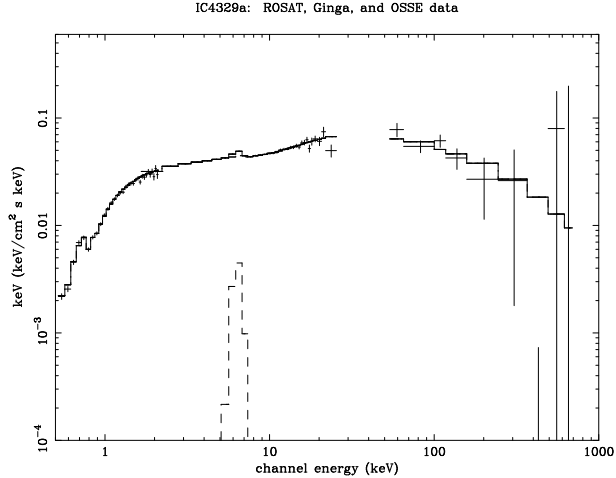}}
\vspace{5pt}
\caption{
Broad-band high energy spectrum of bright Seyfert~1 galaxy IC~4329a.  
The ROSAT (0.1 - 2 keV) and OSSE (50 - 500 keV) data are simultaneous;
the non-simultaneous Ginga (2 - 20 keV) data were scaled to match the 
ROSAT flux at 2 keV.  The solid line is the best fit, including the
underlying power law with an index $\alpha = 0.9$, absorbed at the low
energies by a combination of neutral and ionized material and
exponentially cut off at $E_c \sim 200$ keV, plus the two ingredients
of reprocessing measured in the X--ray band:  Compton reflection
component, plus a Fe K line 
(from Madejski et al. 1995).  
}\label{4329}
\vspace{-10pt}
\end{figure}

The region beyond the Fe~K line shows general hardening of the spectrum, 
accompanied by the pseudo-edge due to Fe~K;  this is a signature of 
Compton reflection, as discussed above.  The intensity of this component 
is consistent with the reflector being a semi-infinite plane;  it peaks 
at $\sim 30$ keV, and beyond this, it steepens gradually, most likely due 
to the Compton recoil as well as the Klein-Nishina effects.  Of 
course the underlying power law also can steepen there -- we 
only observe a sum of the primary and reflected spectrum.  The detailed 
fits (spectral decomposition) of the data imply that the primary spectrum 
is consistent with the power law having an exponential cutoff at an 
e-folding energy of $\sim 200$ keV.  The recent simultaneous RXTE and 
OSSE observations of this object generally confirm this picture.  
A very important constraint on any theoretical models for radiation 
processes in active galaxies is the {\sl absence} of strong
annihilation line at 511 keV.  

Is this high energy spectrum of IC~4329a unique?  Given the variability of 
active galaxies, simultaneous observations in X--rays and soft 
$\gamma$--rays are required, but these are sparse;  an alternative approach 
by \cite{zdziarsum} is to co-add many non-simultaneous 
observations.  This in fact produced an average spectrum that is remarkably 
similar to the above picture, and thus any theoretical interpretation of 
the data for IC~4329a is probably valid for Seyfert nuclei in general.  

\subsection*{5.2. The ``Ultra-soft'' Seyfert 1s:  the Question of the ``Soft 
Excess''}

It is important to note that luminous objects similar to IC~4329a are a
majority of Seyfert~1s, but there is one important sub-class of Seyferts
that shows decidedly distinct X--ray spectra, differing from the above
description in soft X--rays, below $\sim 2$ keV.  These are the so-called
``ultra-soft'' Seyferts.  First observed in the HEAO data by S. Pravdo and
collaborators, they were suggested to be a possibly distinct class of
active galaxies by F. Cordova on the basis of the Einstein Observatory
Imaging Proportional Counter data.  The EXOSAT data analyzed by
\cite{arnaud} as well as by \cite{turner} showed that in a number of
active galaxies, the extrapolation of the hard power law towards low
energies (taking properly into consideration the absorption from our own
Galaxy) {\sl underpredicted} the soft X--ray flux, indicating that there
is an additional component of X--ray emission.  This, the so-called ``soft
excess,'' meant that in a given object, the ultra-soft component can
co-exist with the hard power law, implying that active galaxies generally
have two-component power law spectra, where either or both components are
visible.  It is important to note that this component appears in both low-
and high-luminosity sources;  for instance, it was observed in the
Einstein Observatory data in the quasar PG~1211+143 by \cite{elvis}.  For
the sources where both components are present, they intersect at $\sim 1 -
2$ keV (but this may be an observational artifact);  the spectrum of the
soft component is very soft (steep), and can be described as a power law
with an index $\alpha$ $\sim 2$ or even steeper (but a power law is
usually not a unique model), while the hard component has just the
``canonical'' hard Seyfert~1 spectrum with $\alpha$ $\sim 1$.  

In the very
few cases that the variability of both components has been measured, it
appears that the two vary independently, showing no correlation between 
them (see, e.g., the case of Mkn~335 in Turner 1988).  On the other
hand, such a lack of correlated variability may be the result of a 
complex spectral deconvolution procedure, since only a tail of the soft
excess component is observable due to intervening absorption in the
ISM of the host, or our own Galaxy. In
fact, a correlation of UV and EUV fluxes was clearly observed in the 
``soft excess'' Seyfert NGC~5548 (Marshall et al. 1997).  This implies
some coupling of the soft excess component to the UV/X--ray reprocessing
cycle discussed in more detail in Section 6 below.  Similar
correlation has been also detected in that source between the 
UV flux and soft X--ray residuals observed simultaneously in Ginga 
spectra (cf. Magdziarz et al. 1998). Since the EUV observations suggest much
higher variability amplitude than the UV data, the lack of
apparent correlation in fainter sources may be related to either variation
of the cut off energy of the tail of the soft excess, or 
confusion with spectral index variations in the hard component. 

\begin{figure} 
\vspace{10pt}
\centerline{\psfig{file=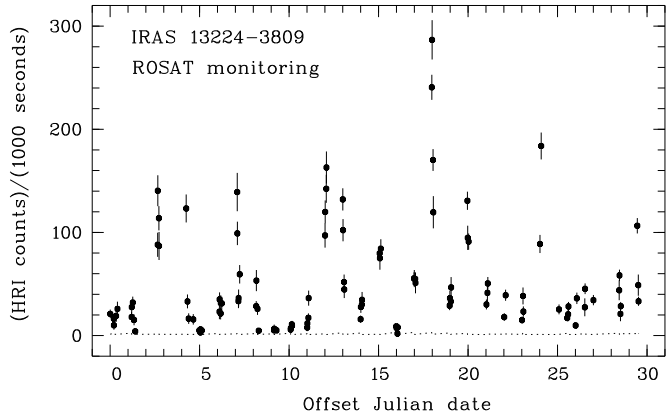,height=2.2in}}
\vspace{10pt}
\caption{
X--ray light curve from the Narrow-Line Seyfert~1 IRAS~13224-3809,
collected with the ROSAT HRI, showing non-linear variability behavior
(from Boller et al. 1997).  Another non-blazar object where non-linearity is
detected is 3C~390.3 (cf. Leighly \& O'Brien 1997)
}\label{iras13224}
\vspace{-5pt}
\end{figure}

``Ultra-soft'' Seyferts appear to be quite common in the ROSAT all-sky
survey, which is not surprising, since ROSAT is a very sensitive
instrument below 2 keV.  Detailed follow-up studies by \cite{boller}
showed a very interesting result:  the optical/UV emission lines in these
objects are generally quite narrow, with the widths on the order of 1,000 -
3,000 km~s$^{-1}$, as compared to $>$ 5,000 km~s$^{-1}$ for ``normal''
Seyfert~1s.  These are {\sl not} similar to Seyfert~2s at all;  these are
widths of permitted lines, while Seyfert~2s only rarely show permitted lines. 
Important clues to the nature of these soft components may be in the
variability patterns, although only very few well-sampled light curves
exist.  The EXOSAT data were consistent with the hard X--ray variability
that is aperiodic but linear, meaning that the time series can be
described as uncorrelated noise (cf. Czerny \& Lehto 1997);  
an example of a light curve for this is
shown in Fig.~1.  The variability of soft X--rays, on the other hand,
is often ``episodic,'' with large flares (up to a factor of 100!) (see 
Fig.~8 for an example), and decidedly non-linear (cf. Boller et
al. 1997).  (In this case a ``non-linear'' behavior manifests itself
qualitatively in an episodic, flare-like behavior such as that illustrated in
Fig.~8;  see, e.g., Vio et al. 1992.  Quantitatively, a ``non-linear'' 
time series is a positive, definite one, which has the ratio of its
standard deviation to its mean which is larger than unity;  see, e.g., 
Green 1993.)  This difference may imply different emission mechanisms in 
the hard and soft X--ray components;  however two observational effects have 
to be considered before drawing any conclusions.  First, the observational
noise or/and presence of additional higher frequency variability in the
hard component may effectively dissolve apparent signatures of
non-linearity (cf. \cite{leighly}). Second, if the variability related
to the energy reprocessing does indeed originate from variations of 
the soft excess component (e.g., Magdziarz et al. 1998), then the 
signatures of non-linearity should be suppressed in the UV and the 
hard continuum. 

Early modeling attempted to describe the ``soft excess'' as the tail end
of the thermal, multi-blackbody emission from an accretion disk;  however,
this seems {\sl not} to be the case, at least for the bright and 
well-studied NGC~5548. Magdziarz et al. (1998) have shown that the UV
component in that object may be associated with rather cold disk
continuum, with a temperature on the order of a few eV, while the soft 
excess requires a separate spectral component.  We will
return to this below.  


\subsection*{5.3. High Energy Spectra of Seyfert 2s vs. Seyfert 1s}

As it was mentioned above, the popular ``unification'' picture explains the 
differences between the spectra of Seyfert~1s vs. Seyfert~2s as due to the 
orientation effects.  To the first order, the only difference that should 
be seen in the X--ray spectra of the two classes is the amount of 
photoelectric absorption, while the underlying continuum should be the 
same.  Just as in the case of Seyfert~1s, this requires simultaneous 
observations by multiple satellites, only the problem is more acute here, 
as the large amount of absorption leaves generally fewer soft X--ray 
photons to allow for a sensitive measurement of the continuum.  Nonetheless, 
the observations with the Ginga satellite by \cite {awaki} 
revealed that X--ray 
spectra of Seyfert~2s are in fact equivalent to spectra of Seyfert~1s, 
absorbed by various column densities of cold gas.  More detailed studies by 
\cite{smith} implied that the continua of Seyfert~2s may be 
somewhat harder, but only marginally so;  a more conclusive results should 
be obtained from observations by the Rossi X--ray Timing Explorer, which 
features a broader bandpass, extending to 50 keV (or, for brighter sources, 
even to 100 keV).  

\begin{figure} 
\vspace{-20pt}
\centerline{\psfig{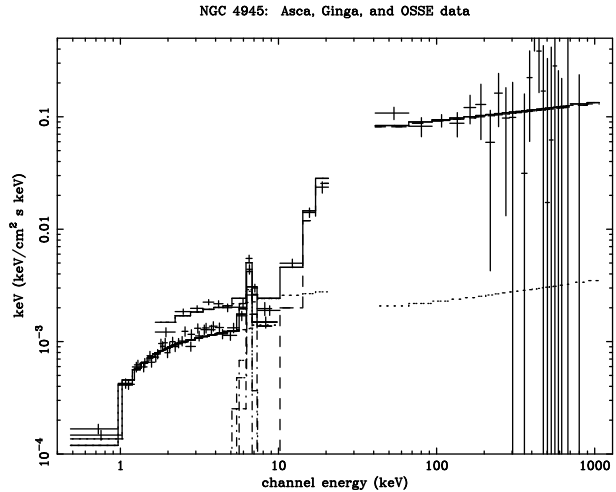}}
\vspace{2pt}
\caption{
Broad-band high energy spectrum of hard X--ray - bright but heavily
absorbed Seyfert~2 galaxy NGC~4945.  
The Asca (0.5 - 10 keV), Ginga (2 - 20 keV) and OSSE (50 - 500 keV)
data are not simultaneous, and thus the overall spectrum may be 
inaccurate as a result of possible source variability.  Nonetheless, the
heavily absorbed primary continuum above $\sim 10$ keV 
is clearly discernible from the scattered / diffuse component below 
$\sim 10$ keV
(from Done, Madejski, \& Smith 1996).  
}\label{4945}
\end{figure}

In any case, the inferred column densities in Seyfert~2s are 
$\sim 10^{22}$ cm$^{-2}$ or more.  In a few cases -- as, for instance, 
the well-studied NGC~1068 -- we only know that no primary X--ray continuum 
is seen, so the absorber must be quite Thomson-thick, with the absorbing 
column greater than $\sim 10^{25}$ cm$^{-2}$.  A good, illustrative example 
of a nearly-extreme Seyfert~2 -- but, with the primary continuum still 
barely penetrating the absorber -- is NGC~4945, a Seyfert~2 which also 
shows megamaser emission, and thus is most likely observed in the plane of 
the putative torus.  In the Asca and Ginga ranges (below $\sim 10$ keV), 
the source is relatively faint, but above $\sim 10$ keV, the spectrum rises 
sharply (\cite{iwasawa4945}).  OSSE observations of it by \cite{done} revealed 
that at 50 - 100 keV, this is the second brightest radio-quiet active 
galaxy in the sky (see Fig.~9).  The absorbing column is large, $\sim 4 
\times 10^{24}$ cm$^{-2}$, and while the observations were not simultaneous, 
a comparison of Fig.~7 and Fig.~9 reveals that the underlying continuum, 
to the first order, is consistent with that of a Seyfert~1.  Another
interesting aspect of this source is the fact that the megamaser
distribution implies a mass of the black hole of $\sim 10^6$ M$_{\rm o}$
(Greenhill et al. 1997).  With the bolometric luminosity of the nucleus
of at least $\sim 10^{42}$ erg s$^{-1}$ (Iwasawa et al. 1993;  Done et
al. 1996), this source 
radiates at a few percent of Eddington luminosity, and thus is
unlikely to be advection-dominated, as may be the case for NGC~4258.  

\subsection*{5.4. High Energy Spectra of High Luminosity Sources}

So far, we discussed primarily the relatively low-luminosity quasars, and 
an obvious question to be asked is:  does this general picture hold for the 
higher luminosity counterparts?  In general, the answer is yes, but with 
some modifications.  It is important here to compare ``apples to apples,'' 
and in the case of quasars, this means selecting radio-quiet objects, as 
radio-loud quasars tend to have higher X--ray -- to -- optical flux ratios, 
which may be due to a contamination by a possible jet;  unfortunately, 
this paucity of X--ray photons in the radio-quiet objects makes spectral 
studies somewhat more difficult.  In general, the more luminous 
objects show a lower ratio of X--ray to optical luminosities 
(see, e.g., Kriss \& Canizares 1985;  Avni \& Tananbaum 1986).  
The recent work by Laor et al. (1997) (using the ROSAT PSPC data) 
and \cite {nandraquas} (using Asca data) indicates that the continuum X--ray 
spectra of higher luminosity objects appear generally similar to those
of the lower luminosity counterparts.  
Notable exception is an absence of the Fe K line and the Compton reflection 
component in quasars (see, e.g., Nandra et al. 1997b).  
With the more luminous central source, this may be the result 
of a nearly-complete ionization of the accreting material, such that 
the reflection component is present, but cannot be distinguished by its 
tell-tale Fe K line and spectral hardening above $\sim 8$ keV:  for an 
entirely ionized reflector, the incident or emerging photons encounter no 
photoelectric absorption, and the reflection is (below $\sim 30$ keV) 
identical to the incident spectrum, with no Fe K line present.  
Unfortunately, those quasars are generally too faint to be studied in 
detail above $\sim 10$ keV, where the only instruments currently available 
are non-imaging proportional counters such as the RXTE, dominated 
by uncertainties of the instrumental as well as the Cosmic X--ray Background.
As a result, any detailed studies must await X--ray reflective optics 
sensitive beyond 10 keV, already under development;  such telescopes 
are essential for studies of these faint objects, as they permit subtraction 
of background from the same image as the source, and thus will yield the 
best quality data for luminous quasars.  

\section*{6. Radiation Processes in Radio-Quiet Active Galaxies}

The availability of good quality high energy spectra permits us to
constrain the possible emission mechanisms that can operate in active
galaxies.  We briefly discussed the case of blazars above; 
observationally, the continuum high energy spectra of radio-quiet 
objects are also decidedly non-thermal, but 
these mechanisms are probably somewhat different than the synchrotron +
Compton model discussed for the blazar jets.  The early work by
\cite{suntit}, using the diffusion approximation, suggested that a
power-law spectrum can be produced by a repeated Compton-upscattering of
soft photons by a Compton-thick bath of hot electrons.  A more general
variant of this model is essentially what is used today to explain the
primary, high energy spectrum in these objects as well as in the Galactic
black hole candidates.  In summary, a successful model has to explain a
power law spectrum with an energy index $\alpha$ of $\sim$ 0.9 -- 1, 
exponentially cutting off at $E_{\rm c} \sim 200$
keV:  other spectral features are merely signatures of reprocessing. 

There are essentially two flavors of the Comptonization model that can be
applied to the isotropic emission in quasars: the original thermal
Comptonization version, and a non-thermal variant.  The difference between
the two is related to the distribution of electron energies.  The
non-thermal version, developed by A. Zdziarski, A. Lightman, P. Coppi, as
well as by C. Done, R. Svensson, G. Ghisellini, and A. Fabian (Zdziarski
et al. 1990;  Zdziarski \& Coppi 1991) involves acceleration of particles
to relativistic energies, and a subsequent pair cascade;  these particles
Comptonize UV photons, believed to be produced in abundance by the
accretion disk.  The attractive feature of this model is that the pairs
thermalize to relatively low temperatures ($\sim$ a few keV), providing
the medium which, again, upscatters the UV photons to form the ``soft
excess'' discussed above. 

One feature, however, predicted by this version of the model, is the
annihilation line that should be present at $\sim 511$ keV.  No 
spectrum of any active galaxy collected so far with the CGRO OSSE 
detector showed such a
feature (Johnson et al.\ 1997), and thus the model is somewhat out of
favor, despite its natural ability to provide the ``soft excess.''
Refinements to the thermal Comptonization model, primarily by \cite{pout},
allowed the quasi-analytical calculation in the optically thin - to
intermediate regime.  This is the most viable current model for
radio-quiet active galaxies, and is discussed in more detail in Chapter
5.3 by J. Poutanen.  Roughly, for regime relevant to the hard X--ray 
spectra of Seyferts, the energy of the 
exponential cutoff $E_{c}$ determines the temperature of the Comptonizing
plasma, such that $kT_{\rm plasma}$ $\simeq$ $E_{c}/1.6$, 
while the index of the power law, together with the cutoff,
determine its optical depth $\tau_{\rm Th}$, such that $\tau_{\rm Th}$ 
$\simeq$ $0.16$ / $(\alpha \times (kT_{\rm plasma}/m_e c^2))$ 
(Pietrini \& Krolik 1995; Poutanen, Krolik, \& Ryde 1997).  
For Seyferts, the Comptonizing plasma has to have a
temperature of $\sim 100$ keV, and optical depth $\tau \sim$ 1;  the
soft ``seed'' photons are available in abundance from the inner accretion
disk, as evidenced by the presence of the reflection component (Zdziarski
et al. 1997).  However, a simple ``sandwich'' type structure (with a 
cold disk covered by a uniform corona) cannot
work, as this would produce too many ``seed'' photons for
Comptonization.  Instead, an example of a 
good phenomenological model is a ``patchy
corona'' above a surface of the disk, proposed by Haardt, Maraschi, \&
Ghisellini (1994);  however,
none of these models address the processes responsible for the particle
acceleration. 

The issue of the ``soft excess'' remains unresolved in the context of the
above models.  Spectral fitting to the data for the well-studied NGC~5548
imply that the soft excess can be produced by a relatively cold ($kT \sim
200$ eV) but optically thick ($\tau > 10$) plasma, while the hard continuum
requires $kT \sim 50$ keV, but $\tau \sim 2$ (Magdziarz et al. 1998).
Coexistence of such two phases may be related to the disk structure and
dynamics of possible multi-phase transition regions (see, e.g.,
Magdziarz \& Blaes 1998).  In fact, such multi-phase medium appears 
naturally in local solutions of the disk corona transition layer 
(see, e. g., \cite{rozanska}).  However, without high quality 
observational data on the spectral shape and temporal 
correlation between
both components, so far, we lack clear clues as to the time evolution
of the plasma energetics.  The author's prejudice (based partially on the
``episodic'' nature of the soft light curves) is that we probably witness
some form of a limit cycle operating in the inner disk, and thus the best
avenue for this is a development of a more detailed theory for the
structure of the inner accretion region, and in particular, the issue of
stability of the transition region between the {\sl bona fide} disk and
the matter free-falling onto the black hole.  However, any tests of
theories require sensitive observations:  especially needed are
well-sampled light curves obtained simultaneously over a broad energy
range, from the softest energies accessible ($\sim 0.1$ keV) up to the end
of the observable spectrum, in the MeV range.  The prospects for such
observations are very good:  with the impending launch of AXAF, XMM,
Astro-E, Integral, and, eventually, Constellation-X, 
we should have the data for the more definitive modeling. 

\vskip 0.2 cm

\noindent {\bf Acknowledgements:}  The author wishes to acknowledge
helpful comments from Drs. J. Krolik, P. Magdziarz, C. Done, E. Boldt, 
and M. Sikora, and figures from Drs. W. Brandt, K. Nandra, and L. Greenhill.


\begin{thebibliography}{}

\bibitem[Abramowicz et al. 1995]
{abram95}
Abramowicz, M., Chen, X., Kato, S., Lasota, J. P., \& Regev, O. 1995,
ApJ, 438, L37

\bibitem[Antonucci \& Miller (1985)]
{anton}
Antonucci, R. R. J, \& Miller, J. S. 1985, ApJ, 297, 621

\bibitem[Arnaud et al. (1985)]
{arnaud}
Arnaud, K. A., et al. 1985, MNRAS, 217, 105

\bibitem[Avni \& Tananbaum 1986]
{avni}
Avni, Y., \& Tananbaum, H. 1986, ApJ, 305, 83

\bibitem[Awaki et al. (1991)]
{awaki}
Awaki, H., Koyama, K, Inoue, H., \& Halpern, J. 1991, PASJ, 43, 195

\bibitem[Blandford \& Levinson (1995)]
{blandlev}
Blandford, R. D., \& Levinson, A. 1995, ApJ, 441, 79

\bibitem[Blandford \& Znajek (1977)]
{blandznajek}
Blandford, R. D., \& Znajek, R. L. 1977, MNRAS, 179, 433

\bibitem[Boller, Brandt, \& Fink (1996)]
{boller}
Boller, T., Brandt, W. N., \& Fink, H. 1996, A\&A, 305, 53

\bibitem[Boller et al. (1997)]
{boll97}
Boller, T., Brandt, W. N., Fabian, A. C., \& Fink, H. 1997, MNRAS, 289, 393

\bibitem[Czerny \& Lehto 1997]
{czernylehto}
Czerny, B., \& Lehto, H. 1997, MNRAS, 285, 365

\bibitem[Dermer, Schlikheiser, \& Mastichiadis (1992)]
{dermer}
Dermer, C., Schlikheiser, R., \& Mastichiadis, A. 1992, A\&A, 256, L27


\bibitem[Done, Madejski, \& Smith (1996)]
{done}
Done, C., Madejski, G. M., \& Smith, D. 1996, ApJ, 463, L63

\bibitem[Eckart \& Genzel (1997)]
{genzel} 
Eckart, A., \& Genzel, R. 1997, MNRAS, 284, 576

\bibitem[Elvis, Wilkes, \& Tananbaum (1985)]
{elvis}
Elvis, M., Wilkes, B., \& Tananbaum, H. 1985, ApJ, 292, 357

\bibitem[Fabian et al. (1993)]
{fabian}
Fabian, A., Nandra, K., Celotti, A., Rees, M., Grove, E., \&
Johnson, W. 1993, ApJ, 416, L57

\bibitem[Ford et al. (1994)]
{ford87}
Ford, H. C., et al. 1994, ApJ, 435, L27

\bibitem[Ford et al. (1998)]
{ford}
Ford, H. C., Tsvetanov, Z. I., Ferrarese, L., \& Jaffe, W. 1998, in
{\sl The Central Regions of the Galaxy and Galaxies}, proc. IAU
Symp. 184, in press

\bibitem[George \& Fabian (1991)]
{george}
George, I. M., \& Fabian, A. C. 1991, MNRAS, 249, 352

\bibitem[Green 1993]
{green93}
Green, A. 1993, PhD Thesis, University of Southampton, UK

\bibitem[Greenhill (1998)]
{greenhill1068}
Greenhill, L. J. 1998, in {\sl Radio Emission from Galactic and Extragalactic
Compact Sources}, proc. IAU Coll. 164, eds. J. Zensus et al., 
ASP Conference Series, in press.  

\bibitem[Greenhill, Moran, \& Herrnstein (1997)]
{greenhill4945}
Greenhill, L. J., Moran, J. M., \& Herrnstein, J. R. 1997, ApJ, 481, L23

\bibitem[Haardt, Maraschi, \& Ghisellini (1994)]
{haardt}
Haardt, F., Maraschi, L., \& Ghisellini, G. 1994, ApJ, 432, L95

\bibitem[Halpern (1982)]
{halp}
Halpern, J. 1982, PhD Thesis, Harvard University

\bibitem[Harms et al. (1994)]
{harms}
Harms, R., et al. 1994, ApJ, 435, L35

\bibitem[Ho et al. (1998)]
{ho}
Ho, L. C., 1998, in {\sl The Central Regions of the Galaxy and 
Galaxies}, proc. IAU Symp. 184, in press

\bibitem[Ichimaru 1987]
{ichimaru}
Ichimaru, S. 1987, ApJ, 214, 840

\bibitem[Iwasawa et al. 1993]
{iwasawa4945}
Iwasawa, K., et al. 1993, ApJ, 409, 155

\bibitem[Iwasawa et al. (1998)]
{iwasawa}
Iwasawa, K., Fabian A. C., Brandt W. N., Kunieda H., Misaki K., 
Reynolds C. S., \& Terashima Y. 1998, MNRAS, submitted

\bibitem[Johnson et al. (1997)]
{johnson}
Johnson, W. N., Zdziarski, A. A., Madejski, G. M., Paciesas, W. S.,
Steinle, H., \& Lin, Y.-C. 1997, in {\sl Proceedings of the Fourth
Compton Symposium}, eds. C. D. Dermer, M. S. Strickman, \& J. D. Kurfess
(AIP:  New York), AIP Conference Proceedings 410, p. 283

\bibitem[K\"onigl (1981)]
{konigl} 
K\"onigl, A. 1981, ApJ, 243, 700

\bibitem[Kormendy \& Richstone (1995)]
{korm}
Kormendy, J., \& Richstone, D. 1995, ARA\&A, 33, 581

\bibitem[Kriss \& Canizares 1985]
{kriss}
Kriss, G. A., \& Canizares, C. 1985, ApJ, 297, 177

\bibitem[Krolik, Done, \& Madejski (1993)]
{krolik}
Krolik, J., Done, C., \& Madejski, G. 1993, ApJ, 402, 432

\bibitem[Laor et al. (1997)]
{laor}
Laor, A., Fiore, F., Elvis, M., Wilkes, B., \& McDowell, J. 1997, 
ApJ, 477, 93

\bibitem[Lasota et al. (1996)]
{lasota}
Lasota, J.-P., Abramowicz, M., Chen, X., Krolik, J., Narayan, R., \&
Yi, I. 1996, ApJ, 462, 142

\bibitem[Leighly \& O'Brien 1997]
{leighly}
Leighly, K., \& O'Brien, P. 1997, ApJ, 481, L15

\bibitem[Lightman \& White (1988)]
{light}
Lightman, A. P., \& White, T. 1988, ApJ, 335, 57

\bibitem[Madejski et al. (1995)]
{mad}
Madejski, G. M., et al. 1995, ApJ, 438, 672

\bibitem[Madejski et al. (1993)]
{6814}
Madejski, G. M., et al. 1993, Nature, 365, 626

\bibitem[Magdziarz et al. (1998)]
{magdziarz}
Magdziarz, P., Blaes, O., Zdziarski, A., Johnson, W., \& 
Smith, D. 1998, MNRAS, in press

\bibitem[Magdziarz \& Blaes (1998)]
{magdziarz_blaes}
Magdziarz, P., \& Blaes, O. 1998, in Proc. IAU Symp. 
188, Kyoto, Japan, in press

\bibitem[Makishima (1986)]
{makis}
Makishima, K. 1986, in {\sl The Physics of Accretion onto Compact Objects},
ed. K. Mason, M. Watson, \& N. White (Springer-Verlag:  Berlin), p. 249

\bibitem[Maoz 1995]
{maoz}
Maoz, E. 1995, ApJ, 447, L91

\bibitem[Maraschi et al. (1994)]
{maraschi}
Maraschi, L., et al. 1994, ApJ, 435, L91

\bibitem[Marshall et al. 1997]
{marshall}
Marshall, H. L., et al. 1997, ApJ, 479, 222

\bibitem[McHardy 1989]
{mchardy}
McHardy, I. 1989, in {\sl Two Topics in X--ray Astronomy}, Proc. 23rd 
ESLAB Symp., eds. N. White, J. Hunt \& B. Battrick, (ESA Publications:  
Paris), vol. SP-296, p. 1111

\bibitem[Miyoshi et al. (1995)]
{miyoshi}
Miyoshi, M., et al. 1995, Nature, 373, 127

\bibitem[Moderski, Sikora, \& Lasota (1998)]
{moder}
Moderski, R., Sikora, M., \& Lasota, J.-P. 1997, in {\sl
Relativistic Jets in AGNs}, eds. M. Ostrowski et al. (Astronomical
Observatiory of the Jagiellonian University:  Krakow) p. 110

\bibitem[Mushotzky (1980)]
{mush}
Mushotzky, R. F. 1980, Adv. Sp. Res., 3, 10

\bibitem[Mushotzky, Done, \& Pounds (1993)]
{mushdp}
Mushotzky, R. F., Done, C., \& Pounds, K. A. 1993, ARA\&A, 31, 717

\bibitem[Nandra et al. (1997a)]
{nandraline}
Nandra, K., George, I., Mushotzky, R. F., Turner, T. J., \& Yaqoob,
T. 1997a, ApJ, 477, 602

\bibitem[Nandra et al. (1997b)]
{nandraquas}
Nandra, K., Mushotzky, R. F., George, I., Turner, T. J., \& Yaqoob,
T. 1997b, ApJ, 488, L91

\bibitem[Narayan \& Yi 1994]
{narayan}
Narayan, R., \& Yi, I. 1994, ApJ, 428, L13

\bibitem[Neufeld \& Maloney 1995]
{maloney}
Neufeld, D. A., \& Maloney, P. R. 1995, ApJ, 447, L17

\bibitem[Papadakis \& Lawrence 1995]
{papa4051}
Papadakis, I. E., \& Lawrence, A. 1995, MNRAS, 272, 161

\bibitem[Papadakis \& Lawrence 1993]
{papa5548}
Papadakis, I. E., \& Lawrence, A. 1993, Nature, 361, 250

\bibitem[Pietrini \& Krolik 1995]
{pietkrol}
Pietrini, P., \& Krolik, J. 1995, ApJ, 447, 526

\bibitem[Pounds et al. (1990)]
{pounds}
Pounds, K., Nandra, K., Stewart, G., George, I., \& Fabian,
A. 1990, Nature, 344, 132

\bibitem[Poutanen \& Svensson (1996)]
{pout}
Poutanen, J., \& Svensson, R. 1996, ApJ, 470, 249

\bibitem[Poutanen, Krolik, \& Ryde 1997]
{poutkrol}
Poutanen, J., Krolik, J., \& Ryde, F. 1997, in {\sl Proceedings of the Fourth
Compton Symposium}, eds. C. D. Dermer, M. S. Strickman, \& J. D. Kurfess
(AIP:  New York), AIP Conference Proceedings 410, p. 972

\bibitem[Ptak (1997)]
{ptak}
Ptak, A. 1997, PhD thesis, University of Maryland, College Park, MD

\bibitem[Rees (1967)]
{rees}
Rees, M. J. 1967, MNRAS, 135, 345

\bibitem[Reynolds et al. (1996)]
{reynolds}
Reynolds, C., DiMateo, T., Fabian, A., Hwang, U., \& Canizares, C. 1996, MNRAS,
283, L111

\bibitem[Reynolds \& Begelman (1998)]
{reynbeg}
Reynolds, C., \& Begelman, M. C. 1998, ApJ, in press

\bibitem[Rothschild et al. (1983)]
{roth}
Rothschild, R. E., Mushotzky, R. F., Baity, W. A., Gruber, D. E.,
Matteson, J. L., \& Peterson, L. E. 1983, ApJ, 269, 423

\bibitem[R\'o\.za\'nska 1998]
{rozanska} R\'o\.za\'nska, A., 1998, MNRAS submitted

\bibitem[Salpeter (1964)]
{salp}
Salpeter, E. E. 1964, ApJ, 140, 796

\bibitem[Schmidt (1963)]
{schm}
Schmidt, M. 1963, Nature, 197, 1040

\bibitem[Sikora (1997)]
{sikora}
Sikora, M., 1997, in {\sl Proceedings of the Fourth
Compton Symposium}, eds. C. D. Dermer, M. S. Strickman, \& J. D. Kurfess
(AIP:  New York), AIP Conference Proceedings 410, p. 494

\bibitem[Sikora, Begelman, \& Rees (1994)]
{sikbegrees}
Sikora, M., Begelman, M. C., \& Rees, M. 1994, ApJ, 421, 153

\bibitem[Sikora et al. (1997)]
{sikoraetal}
Sikora, M., Madejski, G. M., Moderski, R., \& Poutanen, J. 1997, ApJ,
484, 108

\bibitem[Smith \& Done (1996)]
{smith}
Smith, D., \& Done, C. 1996, MNRAS, 280, 355

\bibitem[Sunyaev \& Titarchuk (1980)]
{suntit}
Sunyaev, R., \& Titarchuk, L. 1980, A\&A, 86, 121

\bibitem[Tanaka et al. (1995)]
{tanaka}
Tanaka, Y., et al. 1995, Nature, 375, 659

\bibitem[Turner (1988)]
{turner88}
Turner, T. J. 1988, PhD Thesis, University of Leicester, UK

\bibitem[Turner \& Pounds (1989)]
{turner}
Turner, T. J., \& Pounds, K. 1989, MNRAS, 240, 833

\bibitem[Ulrich, Maraschi, \& Urry (1997)]
{urry}
Ulrich, M.-H., Maraschi, L., \& Urry, C. M. 1997, ARA\&A, 35, 445

\bibitem[Vermeulen \& Cohen (1994)]
{verm}
Vermeulen, R. C., \& Cohen, M. H. 1994, ApJ, 430, 467

\bibitem[Vio et al. 1992]
{vio}
Vio, R., Cristiani, S., Lessi, O., \& Provenzale, A. 1992, ApJ, 391, 518

\bibitem[Wilson \& Colbert (1995)]
{wilson}
Wilson, A., \& Colbert, E. 1995, ApJ, 438, 62

\bibitem[Yi 1996]
{yi}
Yi, I. 1996, ApJ, 473, 645

\bibitem[Zdziarski et al. (1990)]
{zdziar}
Zdziarski, A. A., Ghisellini, G., George, I. M., Svensson, R., Fabian,
A. C., \& Done, C. 1990, ApJ, 363, L1

\bibitem[Zdziarski \& Coppi (1991)]
{zdziarcoppi}
Zdziarski, A., \& Coppi, P. 1991, ApJ, 376, 480

\bibitem[Zdziarski et al. (1995)]
{zdziarsum}
Zdziarski, A. A., Johnson, W. N., Done, C., Smith, D., \&
McNaron-Brown, K. 1995, ApJ, 438, L63

\bibitem [Zdziarski et al. 1997]
{zdziar97}
Zdziarski, A. A., Johnson, W. N., Poutanen, J., Magdziarz, P., \&
Gierlinski, M. 1997, in The Transparent Universe, eds. C. Winkler at
al. (ESA:  Paris), SP-382, 373

\bibitem[Zeldovich \& Novikov (1964)]
{zeld}
Zeldovich, Y., \& Novikov, I. 1964, Sov. Phys. Dokl., 158, 811

\bibitem[Zhang, Cui, \& Chen (1997)]
{zhang}
Zhang, S. N., Cui, W., \& Chen, W. 1997, ApJ, 482, L55

\end{thebibliography}
\end{document}